\documentclass[preprint,prb,a4paper,floats,titlepage,fleqn,footinbib,lengthcheck,twocolumn,tightenlines,balancelastpage,10pt,superscriptaddress,showpacs]{revtex4-1}
\usepackage{natbib}
\usepackage{amssymb}
\usepackage{amsfonts}
\usepackage{amsmath}
\usepackage{graphicx}
\usepackage{color}
\usepackage{longtable}
\usepackage{revsymb}
\usepackage{units}
\usepackage{nicefrac}
\usepackage{textcomp}
\usepackage{epsfig}
\usepackage{footmisc}
\usepackage{soul}
\usepackage{ulem}

\ifx\pdfoutput\relax\let\pdfoutput=\undefined\fi
\newcount\msipdfoutput
\ifx\pdfoutput\undefined\else
\ifcase\pdfoutput\else
\ifx\paperwidth\undefined\else
\ifdim\paperheight=0pt\relax\else\pdfpageheight\paperheight\fi
\ifdim\paperwidth=0pt\relax\else\pdfpagewidth\paperwidth\fi
\fi\fi\fi
\begin{document}

\author{H. Takeya}
\affiliation{National Institute for Materials Science,1-2-1 Sengen, Tsukuba, Ibaraki
305-0047, Japan}
\author{M. ElMassalami}
\affiliation{Instituto de Fisica, UFRJ, CxP l68528, 21945-970, Rio de Janeiro, Brazil}
\author{H. S. Amorim}
\affiliation{Instituto de Fisica, UFRJ, CxP l68528, 21945-970, Rio de Janeiro, Brazil}
\author{H. Fujii}
\affiliation{National Institute for Materials Science,1-2-1 Sengen, Tsukuba, Ibaraki
305-0047, Japan}
\author{T. Mochiku}
\affiliation{National Institute for Materials Science,1-2-1 Sengen, Tsukuba, Ibaraki
305-0047, Japan}
\author{Y. Takano}
\affiliation{National Institute for Materials Science,1-2-1 Sengen, Tsukuba, Ibaraki
305-0047, Japan}
\title{On the superconductivity of the \textrm{Li}$_{x}$\textrm{RhB}$_{y}$
compositions}

\begin{abstract}
We observed superconductivity ($T_{c}$ $\simeq$2-3 K) in \textrm{Li}$_{x}$%
\textrm{RhB}$_{y}$ intermetallics wherein $x$ and $y$ vary over a wide
compositional range. The crystal\ structure consists of cubic unit-cell ($a$ 
$\simeq$ 12.1 \AA ) with centro-symmetric space group $Pn\bar{3}n$. A weak
but positive pressure-induced increase of $T_{c}$ was observed. The
correlations between the composition and each of the followings were
followed over a wide range of $x$ and $y$: the unit-cell dimensions, $T_{c}$%
, Sommerfeld coefficient $\gamma$, Debye temperature $\theta_{D}$, and
critical fields \textit{H}$_{c1}$ and \textit{H}$_{c2}$. The thermal
evolution of the electronic specific heat within the superconducting phase
was observed to follow a quadratic-in-$T$ behavior. In addition, a
paramagnetic Meissner Effect (PME) is manifested during a low-field-cooled
magnetization cycle. This manifestation of quadratic-in-$T$ behavior and PME
feature will be discussed.
\end{abstract}

\maketitle

\section{Introduction}

Recently we reported\cite{10-Li-Rh-B-JPSJ}\ the observation of
superconductivity ($T_{c}$ $\approx$ 2 to 3 K at ambient pressure) in a
novel ternary \textrm{Li}$_{x}$\textrm{RhB}$_{y}$ phase. This phase was
found to be stable over a wide range of the Li and B content (namely $%
0.6<x<2,1<y<2$) while maintaining (i) the same cubic unit-cell ($\sim$12.1 
\AA , see Fig. \ref{Fig1-Xrd}), (ii) the same normal-state properties (e.g.
Sommerfeld constant $\gamma\sim$3 mJ/molK$^{2}$ and Debye temperatures $%
\theta_{D}\sim $250K), and\ (iii) the same superconducting properties (e.g. $%
T_{c}\sim$3K, $H_{c1}(0)\sim$65Oe, $H_{c2}(0)\sim$14 kOe). It happened that
there is a difficulty in reconciling its thermodynamic properties with the
reported symmetry of its crystal structure (see below).\cite{10-Li-Rh-B-JPSJ}
In this work we resolve this difficulty by carrying out extensive
structural, magnetic, magnetoresistivity, and thermal characterizations.

Our earlier structural analysis showed that based on the observed extinction
rule [absence of (00\textit{l}) lines for \textit{l}=odd], the possible
space groups was proposed to be either \textsl{P2$_{1}$3} or \textsl{P4$_{2}$%
32}. Given that these groups are characterized by a non-centrosymetric
feature and that the Rh atom has a high $Z$-number, then it was concluded
that Anisotropic Spin-Orbit Coupling, ASOC, effects should be manifested.%
\cite%
{Sigrist91-Uncon-SUC,Bauer05-CePt3Si-SUp-Norm,Sigrist07-noncon-Sup-nonCentroSym,Frigeri04-Inverson-symmetry-SUC}
In general, the presence of such ASOC would lead to characteristic features
such as: (i) A removal of the spin\ degeneracy which, in turn, would lead to
an enhanced normal-state Pauli paramagnetism. (ii) An admixture of
even-parity spin-singlet and odd-parity spin-triplet pairing states which
may cause a manifestation of nodes in the structure of the quasiparticle gap
function. As a results of such ASOC influences, many thermodynamical
properties should exhibit a characteristic thermal evolution:\cite%
{Sigrist91-Uncon-SUC,Bauer05-CePt3Si-SUp-Norm,Sigrist07-noncon-Sup-nonCentroSym,Frigeri04-Inverson-symmetry-SUC}
e.g. (i) the superconducting specific heat, $C_{S}(T<T_{c})$, should
manifest a power-in-$T$ behavior,\cite%
{Sigrist91-Uncon-SUC,Bauer05-CePt3Si-SUp-Norm,Sigrist07-noncon-Sup-nonCentroSym,Frigeri04-Inverson-symmetry-SUC}
(ii) the susceptibility of the superconducting state should be increased,
and (iii) the upper critical field $H_{c2}(0)$ should exceed the Pauli
paramagnetic limit $H_{p}$ $\cong$3$k_{B}T_{c}/\mu_{B}\sqrt{2}$.\cite%
{Frigeri05-Sup-NoInversion-CePt3Si}

Our earlier thermodynamical characterization on \textrm{Li}$_{x}$\textrm{RhB}%
$_{y}$ indicated that while $C_{S}(T<T_{c})$ does exhibit a quadratic-in-$T$
behavior (expected for line nodes), neither the susceptibility nor the
evolution of $H_{c2}(T)$ curve confirm such an unconventional\ character.
One of the objectives of the present work is to address this contradiction.

It is shown below that the present structural analysis, carried out on more
than four dozens of samples, indicate that the space group of these \textrm{%
Li}$_{x}$\textrm{RhB}$_{y}$ compositions is the \textit{centro-symmetric} $Pn%
\bar{3}n$ rather than the earlier reported \textit{non}-\textit{%
centro-symmetric}\ \textsl{P2$_{1}$3} or \textsl{P4$_{2}$32}:\cite%
{10-Li-Rh-B-JPSJ}{\ this remove the contradiction between structural and
thermodynamic properties}. On the other hand, the above-mentioned
contradiction among the thermodynamic properties will be discussed in term
of conventional (rather than nonconventional) influences. Finally, we
followed the evolution of the superconducting properties with the variation
in the hydrostatic pressure as well as in the structural and material
properties of these compositions (e.g. the unit-cell volume, stoichiometry,
sample purity, defects concentrations ..etc.).

\section{Experimental}

To the best of our knowledge, the stabilization of ternary Li-Rh-B compound
was reported only for the following nonsuperconducting cases: (i) the
hexagonal \textrm{Li}$_{2}$\textrm{RhB}$_{2}$ ($a$ =8.45 \AA\ and $c$ =4.287 
\AA ),\cite{Jung76-Li2RhB2} and (ii) the orthorhombic Li$_{2}$Rh$_{3}$B$_{2}$
($a$ =5.7712 \AA , $b$ =9.4377 \AA , $c$ =2.8301 \AA ).\cite%
{Baileya07-li2Rh3B2} Badica \textit{et al.}\cite{Badica07-Li2Z3B} attempted
a synthesis of Li$_{2}$Rh$_{3}$B but the product was found to be
multi-phasic consisting mainly of binary boride and elemental Li and B.

Polycrystalline samples of various \textrm{Li}$_{x}$\textrm{RhB}$_{y}$
compositions (0.4 $\leq x\leq$ 3 and 1 $\leq y\leq$ 2) were synthesized by
standard solid state reaction of pure Li lump (99.9 \%), Rh powder (99.95
\%), and crystalline B powder (99 \%). Rh and B were, first, mixed and
pressed into pellets and afterwards, together with Li lump, were placed in a
Ta foil or a BN crucible and sealed in a stainless container under an argon
atmosphere. The container was heated up to 700-900 $^{\circ}$C for 20 h and
followed by furnace cooling. Afterwards these products were annealed at the
same temperature range.

\begin{table*}[tbp] \centering%
\caption{Some parameters of Li$_{x}$RhB$_{y}$ compositions where $x$ and $y$ represent the measured
content of Li and B relative to Rh as determined by ICP method; $a$ is the parameter of the cubic cell (standard deviation reflects  the variation associated with differing sample batches); the onset $T_{c}$ as determined from
magnetization, resistivity, or specific heat curves; $H_{c1}(0$ K$)$ as determined from the magnetization
curves; $H_{c2}$(0 K) as determined from the magnetoresistivity curves; $\beta$ , $\gamma$, and $\delta$$_{L}$
coefficients as determined from the specific heat measurements. $\theta$$_{D}$ is estimated to be within the range 240 - 260 K. The coherence length $\xi$(0) and the penetration
depth $\lambda$(0) were calculated from the following Ginzburg-Landau (GL) expressions:
$H_{c2}(t)$ =$\Phi_{0}$/(2$\pi\xi$(0)$^2$(1-$t$)) and $H_{c1}(0)$ =$\Phi_{0}$/[(4$\pi\lambda$(0)$ ^2$ln($\lambda$(0)/$\xi$(0))],
where $\Phi_{0}$ is the flux quantum and $t$=$T/ T_{c}$. The calculated $H_{c2}$(0 K) was determined from the quadratic and WHH expressions (see text).}%
\begin{tabular}{cccccccccccccc}
\hline\hline
Nominal & \multicolumn{2}{c}{measured} & $T_{c}$ & $a$ & $H_{c1}(0)$ & $%
H_{c2}^{quad}(0)$ & $\xi(0)$ & $\lambda(0)$ & $\alpha$ & $H_{c2}^{WHH}(0)$ & 
$\gamma$ & $\beta$ & $\delta_{L}$ \\ \hline
x,y & x & y & K & \AA  & Oe & kOe & nm & nm &  & kOe & mJ/molK$^{2}$ & 
mJ/molK$^{4}$ & J/molK \\ \hline
\textrm{Li}$_{0.8}$\textrm{RhB}$_{1.5}$ & 0.87 & 1.47 & 2.4 & 12.079(1) & 
83.5 & 13.7 & 15.5 & 15.7 & 0.29 & 9.6 & 3.3 & 0.46 & 0.024 \\ 
\textrm{Li}$_{1.0}$\textrm{RhB}$_{1.5}$ & 0.95 & 1.48 & 2.6 & 12.086(7) & 77
& 8.1 & 20.2 & 20.6 & 0.17 & 5.6 & 2.8 & 0.37 & 0.024 \\ 
\textrm{Li}$_{1.2}$\textrm{RhB}$_{1.5}$ & 1.02 & 1.52 & 2.6 & 12.089(9) & 
65.6 & 14.2 & 14.4 & 14.5 & 0.3 & 9.8 & 2.4 & 0.40 & 0.024 \\ \hline\hline
\end{tabular}
\label{TabI}%
\end{table*}%

The weight loss during the heating process was found to be less than 0.2 \%.
This result had been confirmed by the elemental analysis which was conducted
using the Inductively Coupled Plasma (ICP) method on representative samples.
Before the analysis, we used aqua regia first and then K$_{2}$S$_{2}$O$_{7}$
to completely dissolve Rh. The analytical determinations of each element
(given in Table \ref{TabI}) are close to the nominal compositions: an
assuring result considering that both Li and B are light elements and the
former is volatile.

Structural analysis of all investigated polycrystalline samples were carried
out on a monochromatic Cu $K_{\alpha}$ diffractometer equipped with a Si
detector (representative diagrams are shown in Fig. \ref{Fig1-Xrd}).
Magnetization curves were measured on a superconducting quantum interference
device (SQUID) magnetometer. Bulk samples were cut into a cylindrical shape (%
$\phi$0.45 x0.50 cm) and sealed in a gelatin capsule (all handled in an
inert-gas glove box). Pressure-dependent magnetization curves were measured
with a low-temperature hydrostatic micro pressure cell (up to 1 GPa)
operated within a SQUID environment. Daphne oil was used as a
pressure-transmitting fluid while Sn as a manometer. Magnetoresistance
curves were measured on parallelepiped samples of typical 0.11x0.12x0.45 cm$%
^{3}$ dimensions. We used a conventional DC, four-points method (1 mA) in a
home-made probe which was operated within the environment of the
above-mentioned magnetometer. Zero-field specific heat measurements were
carried out on a semi-adiabatic calorimeter operating within the range of
0.5 $<T<$23 K with a precision better than 4\%.

During all experiments, care was exercised so as to avoid air/moisture
exposure since such exposure was found to cause a dimming of metallic luster
and, furthermore, a reduction in both $T_{c}$ and superconducting volume
fraction. As such, samples were usually covered with apiezon N grease and
guarded in an inert atmosphere (for remeasurement, grease was wiped off).

Some samples show double superconducting transitions in the magnetization,
resistivity, or specific heat. Given that $T_{c}$ of these transitions are $%
\sim$ 2 to 3 K and that the measured diffractograms do not exhibit any of
the known superconducting contaminant phases (see caption of Fig. \ref%
{Fig1-Xrd} and text below), then the manifestation of such double
transitions is most probably related to the granular character of these
samples or to an unknown ternary phase.

\section{Results}

\subsection{Structural Analysis}

\begin{figure}[ht]
\centering
\includegraphics[
height=5.6915cm,
width= 7.8969cm
]{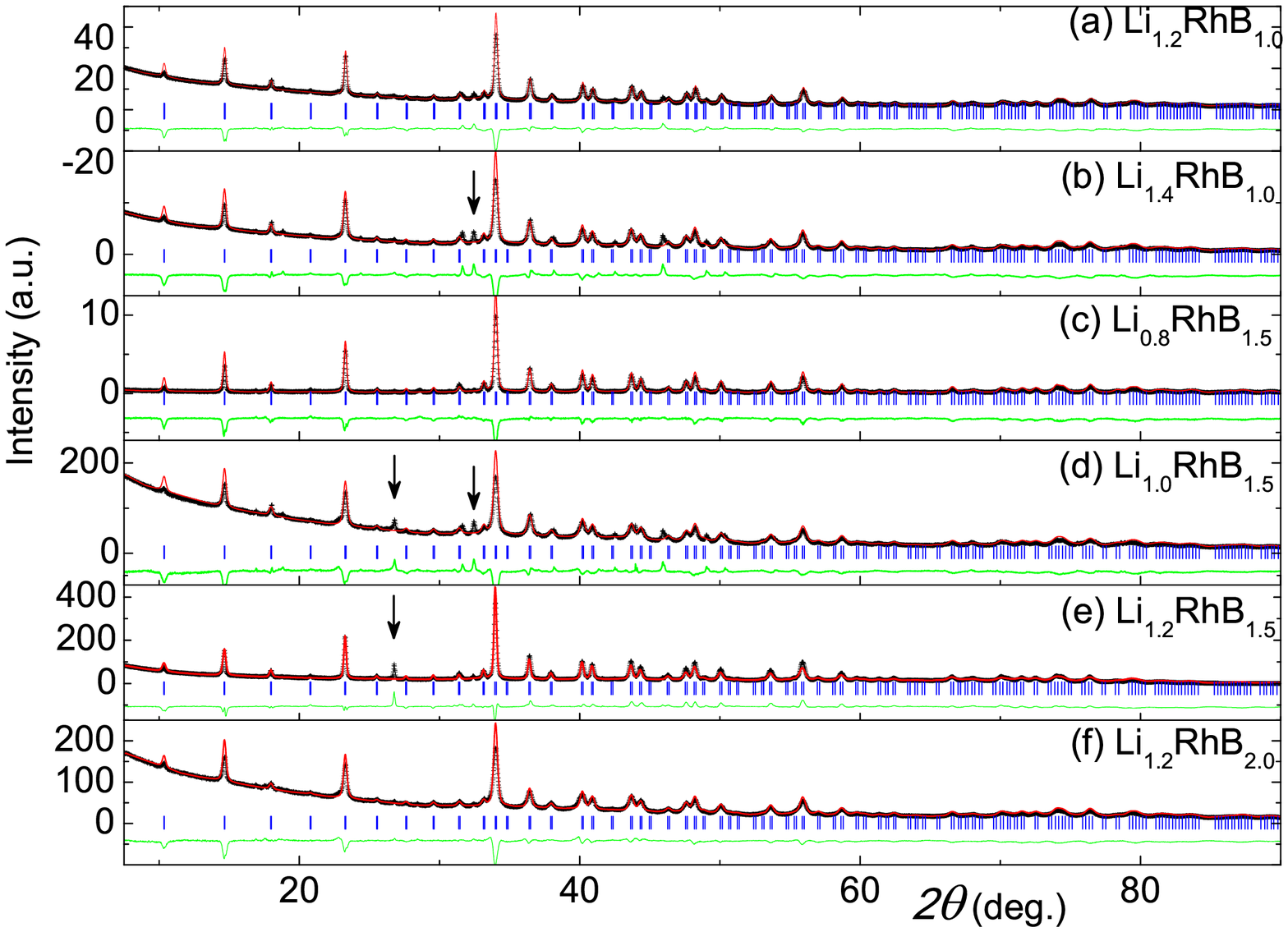}
\caption{(\textit{Color Online})
Representative room-temperature X-ray diffractograms of \textrm{Li}$_{x}$%
\textrm{RhB}$_{y}$\textrm{\ }emphasizing the stability of the almost
single-phase cubic structure in spite of the wide variation in $x$ and $y$. 
\textit{Symbols}: measured intensities; \textit{short bars}: the Bragg
positions, \textit{lower thin line}: the difference curve;\ \textit{solid
line}: Rietveld calculated pattern using \textsl{Pn\={3}n} (see text). The
extremely weak impurity lines (the more intense, marked with short vertical
arrow) are found to be related to nonsuperconducting \textrm{Li}$_{2}$%
\textrm{Rh}$_{3}$\textrm{B}$_{2}$ (more evident at $x>1$and $y\approx 1.5$),
nonsuperconducting \textrm{RhB }(more evident at $x<1$ and $y\approx 1$) and
an unknown tetragonal phase (more evident at $0.4<x<1.5$ and $y\approx 1$).}
\label{Fig1-Xrd}
\end{figure}

\begin{figure}[ht]
\centering
\includegraphics[
height=5.9485cm,
width=8.0309cm
]{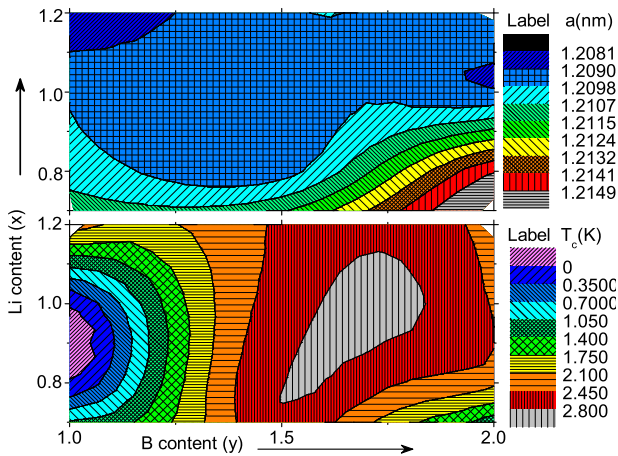}
\caption{(\textit{Color online}) A partial
section of two contour plots representing the evolution of the $a$-parameter
(upper panel) and $T_{c}$ (lower panel) of\textrm{\ Li}$_{x}$\textrm{RhB}$%
_{y}$ with the Li ($x$) and B ($y$) content. These plots are constructed
from compositions with fixed $y$=1, 1.25, 1.5, 1.75, 2 while varying $x$%
=0.4, 0.6, ..., 1.8, 2. It is cautioned that, our lowest measuring
temperature is 1.8 K, accordingly all $T_{c}$ below 1.8 K are generated by
the interpolating plotting subroutine. The observed correlations are
discussed in the text.}
\label{Fig.2-a-T-contour}
\end{figure}

Extensive structural and elemental analyses were carried out on more than
four dozens of \textrm{Li}$_{x}$\textrm{RhB}$_{y}$ compositions covering the
range of 0.6$\leq x\leq$ 1.4 and 0.5$\leq y\leq$ 2 while keeping Rh
stoichiometry fixed. As can be seen in the representatives Figs. \ref%
{Fig1-Xrd}-\ref{Fig.2-a-T-contour}(upper panel) and Table \ref{TabI} as well
as the electron diffraction patterns of Ref. {\onlinecite{10-Li-Rh-B-JPSJ}},
the almost single phase is stable over the range 0.8 $\leq x\leq$ 1.5 and 1 $%
\leq y\leq$ 2 but its concentration is strongly decreased when $x$ or $y$ is
far from this range. All related Bragg lines can be indexed if one adopts a
large-sized cubic unit-cell (see, e.g., Fig. \ref{Fig1-Xrd}).

The wide variation in \textrm{Li}$\ $and \textrm{B} content as well as their
low atomic scattering factors make it extremely difficult to ascertain
correctly their stoichiometry or to calculate the involved density.
Nonetheless, application of extinction rules on resolution-improved
diffractograms suggested that the space group is the centrosymmetric $Pn\bar{%
3}n$. Furthermore,{\ }using Le-Bail method and Patterson maps, the atomic
position of the heavier Rh atoms are conclusively identified (for more
details see Ref.{\onlinecite{LixRhBy-CrystalStructure}}): Rh occupies the
Wyckoff positions $48i$ and $12e$. Furthermore, it seems that B occupies the
position $8c$ leading to the formation of some distorted and others
undistorted octahedrons: a features also common in other Rh-B compositions.
It is noted that although these structural considerations lead only to
partial determination and that further elucidation requires the
identification of the exact stoichiometry and positions of Li and B atoms (a
task which would be much effectively served by using, e.g., neutron
diffraction analysis), nonetheless the calculated Rietveld patterns (Fig. %
\ref{Fig1-Xrd}) compares favorably with experiments and, as such, confirm
the correct identification of the space group, all Rh positions, and one
position of B.\cite{LixRhBy-CrystalStructure} Another encouraging evidence
is that on taking \textrm{Li}$_{x}$\textrm{RhB}$_{y}$ with, say, $x,y\approx$
1 or 1.5, one calculates a density of 6.8 to 7.3 g/cm$^{3}$ which is
comparable to 7.35 g/cm$^{3}$ of \textrm{Li}$_{2}$\textrm{Rh}$_{3}$\textrm{B}%
$_{2}$;\cite{Baileya07-li2Rh3B2} unfortunately, due to the strong porous
character of these materials, we were unable to measure their density by
conventional methods.\cite{LixRhBy-CrystalStructure}

The correlation of the unit-cell $a$-parameter with the Li/B content is
shown in the upper contour plot of Fig. \ref{Fig.2-a-T-contour}: evidently,
on fixing Li\ (B) content and varying B (Li) concentration, the evolution of
the $a$-parameter\ does not reflect any Vegard's law. In fact, a variation
in the small-sized Li and B in \textrm{Li}$_{x}$\textrm{RhB}$_{y}$ over the
whole range of 0.4 $\leq x\leq$ 3 and 1$\leq y\leq$ 2 modifies the unit-cell
volume by only 1.1\%: this emphasizes the crucial role of Rh sublattice.

\subsection{Magnetization}

\begin{figure}[ht]
\centering
\includegraphics[
height=5.5772cm,
width=7.8859cm
]{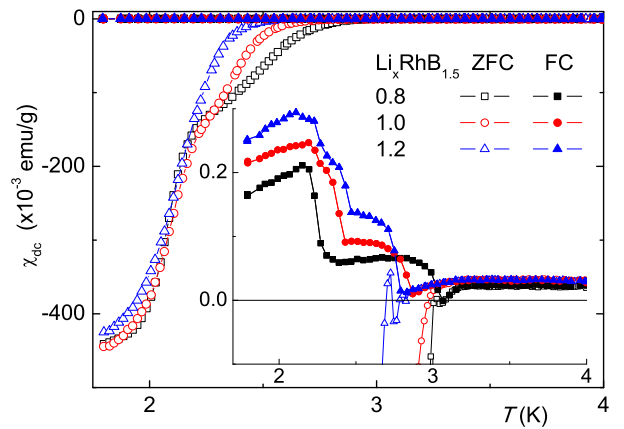}
\caption{(\textit{Color online}) DC
susceptibility curves of Li$_{x}$RhB$_{1.5}$ (x=0.8, 1.0, 1.2) at 20 Oe on
ZFC (open symbols) and\ FC (filled symbols) cycles.\ \textit{Inset}: an
expanded view showing the typical character of the PME:\ on FC, just below $%
T_{c}$, $\protect\chi (T)$ becomes negative and afterwards, on further
cooling, turns into a positive value. By contrast,\ the ZFC susceptibility
exhibits the normally-expected (negative) screening signal. The magnitude of
the PME signal at $H$ = 20 Oe is $\sim $0.1\% of the shielding ZFC signal
(this is reminiscent of the PME in Nb disk\protect\cite%
{Thompson95-ParamMeissner-Nb}). The structure in both ZFC and FC curves
within the immediate neighborhood below $T_{c}$ is related to the fact that,
within this region, the critical currents associated with\ most of the PME
loops are too small to drive spontaneous moments.\protect\cite%
{Khomskii94-ParamagMeissner,Li03-ParamagMeissner,Siegrist-ParamagMeissner-dwave}}
\label{Fig-MvsT}
\end{figure}

\begin{figure}[ht]
\centering
\includegraphics[
height=6.3241cm,
width=8.0309cm
]{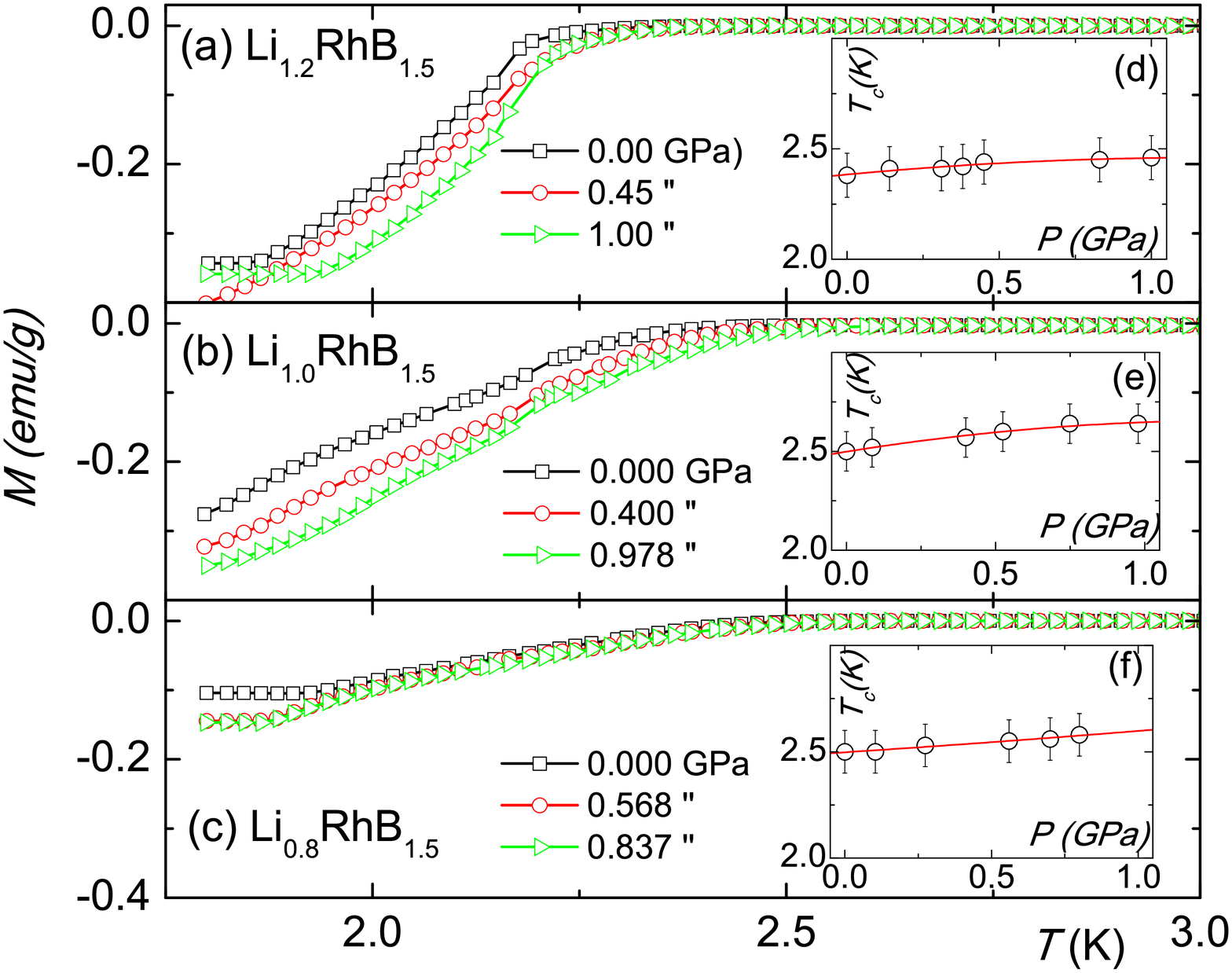}
\caption{(\textit{Color online})
Pressure-dependent magnetization curves of \ (a) Li$_{1.2}$RhB$_{1.5}$, (b)
LiRhB$_{1.5}$, and (c) Li$_{0.8}$RhB$_{1.5}$. \ \textit{Insets}: the
evolution of $T_{c}$ with the applied pressure is approximated as ($p$ in
GPa): (d) $T_{c}\approx 2.38+0.13p-0.05p^{2}$ K for Li$_{1.2}$RhB$_{1.5}$,
(e) $T_{c}\approx 2.50+0.24-0.09p^{2}$ K for Li$_{1.0}$RhB$_{1.5}$ and (f) $%
T_{c}\approx 2.50+0.09+0.01p^{2}$ K for Li$_{0.8}$RhB$_{1.5}$. }
\label{Fig-pressureMvsT}
\end{figure}

Zero-Field-Cooled (ZFC) magnetization of \textrm{Li}$_{x}$\textrm{RhB}$_{y}$
(Fig. \ref{Fig-MvsT}) exhibit a strong shielding signal.\ Field-Cooled (FC)
magnetization, on the other hand, indicates a Paramagnetic Meissner Effect
(PME): a negative drop immediately below $T_{c}$ followed by a surge of a
net (positive) paramagnetism, indicative of spontaneous magnetic moment,
well below $T_{c}$. The XRD diffractograms of these three representative
samples are shown in Fig. \ref{Fig1-Xrd}(c, d, e), each is practically a
single-phase cubic structure. This together with the observed thermal
evolution of this effect (as well as that of isothermal field-dependent
magnetization for $H>H_{c2}$) indicate that it is not related to an
extrinsic contaminating paramagnetic centres. Generally, PME appears in
granular superconductors wherein inverse Josephson Couplings (the so-called $%
\pi $ contacts) are formed at the boundaries of multiply-connected
superconducting grains.\cite%
{Khomskii94-ParamagMeissner,Li03-ParamagMeissner,Siegrist-ParamagMeissner-dwave}
Such boundaries may arise either due to extrinsic (such as disorder or
impurity) or intrinsic factors (such as boundaries that connect differently
oriented crystallites of superconductors with unconventional pairing\cite%
{Siegrist-ParamagMeissner-dwave}). As that the space group is
centrosymmetric then no strong ASOC effects are expected and the presence of
PME is most probably related to extrinsic factors or extreme porosity:\cite%
{LixRhBy-CrystalStructure} indeed the strength of the PME varies within
different batches of the same sample.

\begin{figure}[ht]
\centering
\includegraphics[
height= 6.4844cm,
width=8.0309cm
]{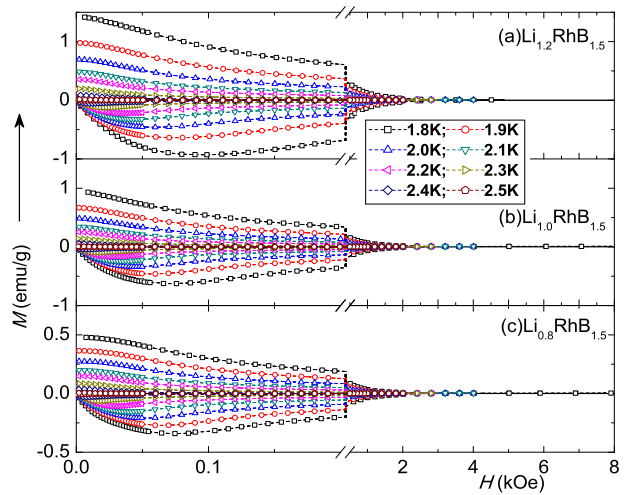}
\caption{(\textit{Color Online}) Isotherm $M(H)$ curves of\ Li$_{x}$RhB$_{1.5}$:
(a) $x$=1.2, (b) $x$=1.0, (c) $x$=0.8. For clarity, the ordinate scale of
the lower panel was expanded by a factor of two.}
\label{Fig-MvsH}
\end{figure}

\begin{figure}[ht]
\includegraphics[
height= 6.0978cm,
width=8.0309cm
]{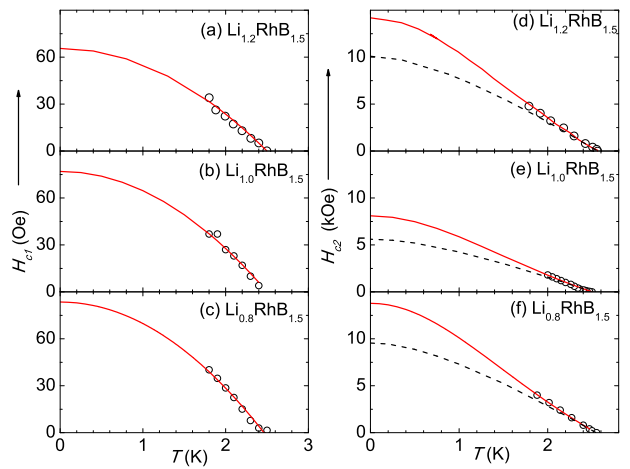}
\caption{(\textit{Color Online}) (a, b, c):\
Measured $H_{c1}(T)$ (symbols) of Li$_{x}$RhB$_{1.5}$ were fitted (solid
lines) to the relation $H_{c1}=H_{c1}(0)(1-t^{2})$, where $t$ =$T/T_{c}$.
(d, e, f): measured $H_{c2}(T)$ of Li$_{x}$RhB$_{1.5}$ (symbol) were
analyzed using (i) the quadratic formula $H_{c2}$($t$) = $H_{c2}$[(1-$t^{2}$%
)/(1+$t^{2}$)] (solid line) and (ii) the WHH expression (dashed lines).}
\label{Fig-Hc1-Hc2}
\end{figure}

\begin{figure}[ht]
\includegraphics[
height=5.955cm,
width=8.0309cm
]{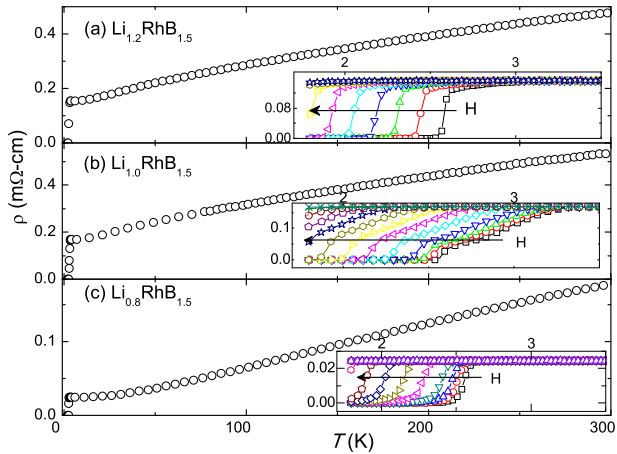}
\caption{Zero-field resistivity versus
temperature of (a) Li$_{1.2}$RhB$_{1.5}$, (b) Li$_{1}$RhB$_{1.5}$, and (c) Li%
$_{0.8}$RhB$_{1.5}$ samples showing \ a normal-state metallic character
above $T_{c}$. Insets: the magnetoresstivity across the superconducting
transition region. For clarity, the ordinate scale of the lower panel was
expanded by a factor of two.}
\label{Fig-RvsT}
\end{figure}

The obtained $T_{c}$ is shown as a function of the Li/B content in Fig. \ref%
{Fig.2-a-T-contour}: $T_{c}$ reaches a maximum within an approximate
triangle-shape region having ($x,y$) vertices as (0.8, 1.5), (1.0, 1.8) and
(1.2, 1.6). Accordingly, the three \textrm{Li}$_{x}$\textrm{RhB}$_{1.5}$ ($%
x=0.8,1.0,1.2$) samples were extensively studied since they are faithful
representatives of the whole series. Their $a$-parameters stabilize around
12.08-12.10 \AA\ (see Fig. \ref{Fig.2-a-T-contour}). Furthermore, a
variation in their Li/B content influences $T_{c}$, most probably through an
induced variation in $N$($E_{F}$), Debye temperature $\theta_{D}$, or
pairing interaction $U$ [e.g. as in the BCS relation $T_{c}=0.85\theta_{D}%
\exp\left( -1/UN(E_{F})\right) $]. As most of these parameters can be varied
through pressure, we investigated as well the influence of applied pressure (%
$p$) on the superconductivity of these \textrm{Li}$_{x}$\textrm{RhB}$_{y}$
compositions. The pressure-dependent magnetization (Fig. \ref%
{Fig-pressureMvsT}) indicates that both the superconducting fraction and $%
T_{c}$ are weakly enhanced. In particular, for pressure up to 1 GPa and to a
second order in $p$, $T_{c}\approx T_{c0}+d_{1}p+d_{2}p^{2}$ ($p$ in GPa)
where $T_{c0}$= 2.50, 2.50, 2.38 K; $d_{1}=\frac{\partial T_{c}}{\partial p}$%
=0.09, 0.24, 0.13 K/GPa; $d_{2}=\frac{\partial^{2}T_{c}}{2\partial p^{2}}$=
0.01,-0.09, -0.05 for \textrm{Li}$_{x}$\textrm{RhB}$_{1.5}$ ($x$%
=0.8,1.0,1.2, resp.). Evidently the overall pressure influence is almost
linear and rather weak. Relatively, \textrm{Li}$_{1.0}$\textrm{RhB}$_{1.5}$
(mid-panel of Fig. \ref{Fig-pressureMvsT}) exhibits a more pronounced $P$%
-induced variation.

Figure \ref{Fig-MvsH} shows the isothermal magnetization curves of \textrm{Li%
}$_{x}$\textrm{RhB}$_{1.5}$: typical type-II curves with no strong positive
normal-state paramagnetic susceptibility (absence of strong polarization).
This latter result is in agreement with the same features exhibited in Figs. %
\ref{Fig-MvsT}-\ref{Fig-pressureMvsT}: all confirm the absence of ASOC
effects. Based on these isothermal curves of Fig. \ref{Fig-MvsH}, we
determined the thermal evolution of $H_{c1}$ (for $H_{c2}$, see below)
which, as can be seen in Figs. \ref{Fig-Hc1-Hc2}(a-c), follows reasonably
well the relation $H_{c1}$=$H_{c1}(0)$[1-($T/T_{c}$)$^{2}$] wherein $%
H_{c1}(0)$ and $T_{c}$ are as given in Table \ref{TabI}.

\subsection{Magnetoresistivity}

The thermal evolution of the resistivity of \textrm{Li}$_{x}$\textrm{RhB}$%
_{1.5}$, $\rho(T_{c}<T\leq300$ K$)$, shown in Fig. \ref{Fig-RvsT}, indicates
a metallic normal state. For most samples, $\rho(T=$300 K) $\leq$ 0.48 m$%
\Omega$-cm while the residual resistivity (in the immediate range above $%
T_{c}$) is $\leq$ 0.12 m$\Omega$-cm: that RRR$\sim$4 suggests additional
scattering processes (e.g. a random atomic distribution, interstitial or
substitutional defects related to Li/B nonstoichiometry). Fig. \ref{Fig-RvsT}
indicate also a superconducting state with a transition which, due to
percolation, are much sharper and narrower than the ones observed in the
magnetizations or specific heats. Because of these advantageous features, $%
H_{c2}(T)$, Fig. \ref{Fig-Hc1-Hc2}, was determined from the midpoint of the
transition occurring in each of $\rho(T,H)$ curve of Fig. \ref{Fig-RvsT}
rather than from the event occurring in each $M(H)$ isotherm of Fig. \ref%
{Fig-MvsH}.

\subsection{Specific Heat}

\begin{figure}[ht]
\includegraphics[
height=5.9089cm,
width= 8.0309cm
]{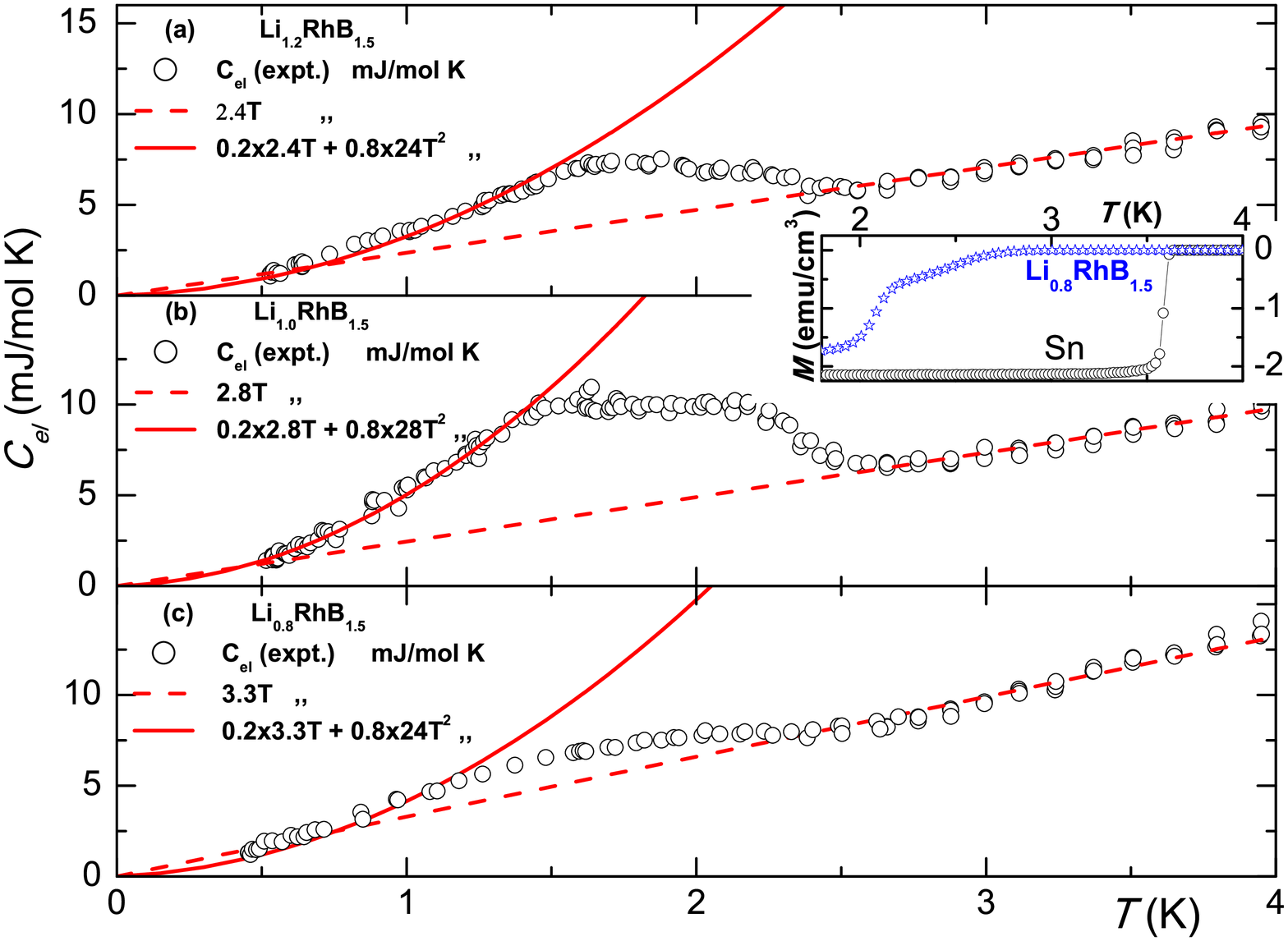}
\caption{(\textit{Color Online}) The
electronic specific heat contribution of \ (a) \textrm{Li}$_{1.2}$\textrm{RB}%
$_{1.5}$, (b) \textrm{Li}$_{1.0}$\textrm{RB}$_{1.5}$ and (c) \textrm{Li}$%
_{0.8}$\textrm{RB}$_{1.5}$ obtained after subtracting the calculated Debye
contribution: $C_{el}=C_{tot}-\protect\beta T^{3}$. The dashed lines are
least-square fits to $C_{el}=\protect\gamma T$ while the solid thick lines
represent the calculation $0.2(\protect\gamma T)+0.80.\protect\alpha%
_{L}\left( T/T_{c}\right) ^{2}$ (see text). The specific heat jump at $T_{c}$
is not sharp but the structure, just below $T_{c}$, is evident as in the
magnetization curve (see inset). \textit{Inset}: As a representative, the
diamagnetism of \textrm{Li}$_{0.8}$\textrm{RB}$_{1.5}$ is 80\% of that of a 
\textrm{Sn} sample having the same shape and size; this is attributed to a
presence of normal-state regions (see text).}
\label{Fig-CbyT-TT}
\end{figure}

The specific heats of \textrm{Li}$_{x}$\textrm{RhB}$_{y}$ samples evolves,
within $T_{c}<T<5$ K, as $\gamma T+\beta T^{3}$ [Ref.{%
\onlinecite{10-Li-Rh-B-JPSJ}}]: $\gamma$ and $\beta$ are given in Table \ref%
{TabI}.\ Just like $T_{c}$ and lattice parameter, there is a weak
Li-dependence of both $\gamma$ and $\beta$; $\gamma$, in particular,
decreases slightly as Li content is increased.

The electronic\ contribution, $C_{el}(T)$, obtained after subtracting the
Debye part ($\beta T^{3}$), is shown in Fig. \ref{Fig-CbyT-TT}. For a
conventional BCS-type gapped superconductor, the electronic contribution
below $T_{c}$ is dominated by an exponential-in-$T$ behavior. Fig. \ref%
{Fig-CbyT-TT} does not show any exponential evolution and as such the gap is
either not fully-developed or blurred by some anomalous behavior. The origin
behind such a behavior may be revealed if one obtains an analytical
expression of the thermal evolution of $C_{el}(T<T_{c})$. But first let us
evaluate whether $C_{el}$ is due solely to the superconducting phase
contribution. In that regard, the inset of Fig. \ref{Fig-CbyT-TT} indicate
that the typical superconducting shielding fraction of these samples is 80\%
of the signal obtained from a similar-sized Sn sample.\cite{10-Li-Rh-B-JPSJ}%
\ It was assumed that the residual normal-state gives rise to a $0.2\gamma T$
contribution. Fig. \ref{Fig-CbyT-TT} indicates that the relation $%
C_{el}\left( T<T_{c}\right) $ = $0.2\gamma T+0.8\delta_{L}\left(
T/T_{c}\right) ^{2}$ describes reasonably well the electronic contribution.
The obtained values of $\delta_{L}$ are similar for all compositions, namely
24 mJ/molK: this suggests a correlation between $\delta_{L}$ and the
electronic contribution of the Rh sublattice since such a contribution does
not vary across the studied samples. Though the arguments given in Ref. {%
\onlinecite{10-Li-Rh-B-JPSJ}} attributed{\ }such a quadratic-in-$T$ behavior
to line nodes, our present understanding\ (based on the above-mentioned
crystallographic and thermodynamics arguments) is that this power relation
may be related to distribution effects arising from, say, variation in the
Li/B content (for evidences regarding $H_{c2}(0)$ see below).

\section{Discussion and Conclusions}

Based on the correlation between $T_{c}$ and the volume (\textit{V}$=a^{3}$)
shown in Fig. \ref{Fig.2-a-T-contour}, one expects:%
\begin{equation*}
\frac{\partial T_{c}}{\partial p}=\frac{1}{\text{\textit{V}}}\frac {\partial%
\text{\textit{V}}}{\partial p}\frac{\text{\textit{V}}\partial T_{c}}{\partial%
\text{\textit{V}}}=\beta.\text{\textit{V}}.\frac{\partial T_{c}}{\partial%
\text{\textit{V}}},
\end{equation*}
where $\beta$ is the compressibility of the solid. Although there are no
information on $\beta(T,p)$,\textit{\ V}$(T,p)$, or lattice anisotropy, it
is possible to correlate the observed weak pressure-dependence of $T_{c}$
with the general features of the upper panel of Fig. \ref{Fig.2-a-T-contour}%
: these particular samples are situated within a \textrm{Li/B} region
wherein the overall variation of $T_{c}$ with \textit{V }(thus with
pressure) is weak; even more weaker dependence is observed for regions with
an excess or deficiency of the Li content. From these features (see also
Table \ref{TabI}), it can be concluded that a variation in pressure or Li
content (B is fixed) would not bring about any strong variation in $T_{c}$, $%
N$($E_{F}$), $\theta_{D}$, or $U$.

The thermal evolution of each $H_{c2}(T)$ curve, shown in Figs. \ref%
{Fig-Hc1-Hc2} (d-f), was analyzed in terms of (i) the quadratic
Ginzburg-Landau relation $H_{c2}$($t$) = $H_{c2}$[(1-$t^{2}$)/(1+$t^{2}$)]
and (ii) the Werthamer--Helfand--Hohenberg (WHH) expression\cite%
{WHH66-Hc2,Hake67-Hc2} which is usually parameterized in terms of $\alpha$
(a measure of the Pauli spin effect) and $\lambda_{so}$ (a measure of the
spin-orbit scattering).\ While $\lambda_{so}$ is a fit parameter, $\alpha$
is taken to be determined experimentally, based on the relation\cite%
{WHH66-Hc2,Hake67-Hc2} $\alpha=5.33\times10^{-5}(\partial H_{c2}/\partial
T)_{T_{c}}$, giving 0.17, 0.29, 0.3 for $x$=0.8, 1.0 and 1.2, respectively.
As expected, both descriptions of $H_{c2}(T)$ reproduce satisfactorily the
measured curves within the range $T\rightarrow T_{c}$. In fact, both
descriptions are reasonable within the available temperature range since
this range is still very close to $T_{c}$ ($t=T/T_{c}\rightarrow1$). In
general, the WHH description is more appropriate for the range $T\rightarrow
0$: accordingly, we used the relation $H_{c2}(0)=-0.693T_{c}(\partial
H_{c2}/\partial T)_{T_{c}}$ to evaluate $H_{c2}(0)$ giving 5.6, 9.6 and 9.8
kOe for $x$=0.8, 1.0 and 1.2, respectively. Such $H_{c2}(0)$ values are
surprisingly low. In fact it is one order of magnitude lower than the
paramagnetic limit $H_{p}$ $\cong$3$k_{B}T_{c}/\mu_{B}\sqrt{2}\simeq$ 80
kOe. As such this constitutes an additional evidence which (together with
the above-mentioned ones) confirms the absence of ASOC\ effects. Indeed, in
spite of the higher $Z$ value of Rh, the WHH analysis of $H_{c2}(T)$ (Fig. %
\ref{Fig-Hc1-Hc2}) indicate no significant role for the $\lambda_{so}$
parameter. Accordingly, $H_{c2}(T)$ in these \textrm{Li}$_{x}$\textrm{RhB}$%
_{y}$ compositions is taken to be determined by the standard orbital driven
depairing process.

In summary, \textrm{Li}$_{x}$\textrm{RhB}$_{y}$ compositions form a new
class of Li-based superconductors. The variation in \textrm{Li/B} ratio is
accompanied by a weak change in the unit-cell length,\ in the normal-state
properties (e.g. $\gamma$,\ $\beta$), and in the superconducting properties
(e.g. $H_{c1}(0)$, $H_{c2}(0)$, and $T_{c}$). Many of the studied parameters
are interrelated: as an example, the $a$-parameter and $T_{c}$ are
correlated and, furthermore, this same correlation is evident in the\
positive pressure dependence of $T_{c}$. As these materials are
centrosymmetric superconductors, the observations of PME during the
field-cooled $M(T)$ cycle and a quadratic-in-$T$ superconducting specific
heat are attributed to conventional (rather than nonconventional) features
such as inhomogeneous distribution of defects or Li/B atoms. Indeed the
thermal evolution of $H_{c2}(0)$\ can be described by a\ conventional WHH
expression. Finally, based on the observed correlation between composition, $%
V$ and $T_{c}$ of \textrm{Li}$_{x}$\textrm{RhB}$_{y}$, it would be very
interesting to carry out a systematic study on the Li-$M$-$X$ series ($M$=
transition metal, $\mathit{X}$ =B, As, Si, Ge).

\begin{acknowledgments}
The authors are grateful to the "Foundation for Promotion of Material
Science and Technology of Japan (MST Foundation)" for the financial support.
We also acknowledge the partial support received from the Japan Society for
the Promotion of Science.
\end{acknowledgments}

\bibliographystyle{apsrev4-1}
\bibliography{addsIin1,ASOCunconvSup,borocarbides,crystalography,intermetallic,massalami,notes,pnictides,SupClassic,swp00000,ToBePublished}

\end{document}